\documentclass[12pt,american,english]{article}
\usepackage[T1]{fontenc}
\usepackage[latin9]{inputenc}
\usepackage{amsmath}
\usepackage{amssymb}
\usepackage{esint}

\makeatletter
\usepackage{a4wide}
\usepackage{ae,aecompl}
\usepackage{bbm}
\usepackage{amsmath}

\makeatother

\usepackage{babel}

\begin{document}
\setcounter{page}{0}

$\,\,$\vspace{1cm}

\begin{center}
\textbf{\LARGE BPS $\mathbb{Z}_{2}$ monopoles and }\\
\textbf{\LARGE ${\cal N}=2$ $SU(n)$ superconformal field theories
}\\
\textbf{\LARGE on the Higgs branch }
\par\end{center}{\LARGE \par}

\vspace{1cm}

\begin{center}
\textbf{Marco A.C. Kneipp}%
\footnote{kneipp@fsc.ufsc.br%
}\textbf{ and Paulo J. Liebgott}%
\footnote{pj.liebgott@gmail.com%
}
\par\end{center}

\begin{center}{\em Departamento de F\'\i sica,\\ 

Universidade Federal de Santa Catarina (UFSC),\\

Campus Universit\'ario, Trindade,\\

88040-900, Florian\'opols, Brazil.  \\ }

\end{center}

\vspace{0.3cm}
\begin{abstract}
We obtain BPS $\mathbb{Z}_{2}$ monopole solutions in Yang-Mills-Higgs
theories with the gauge group $SU(n)$ broken to $Spin(n)/\mathbb{Z}_{2}$
by a{\normalsize{} scalar field in the $n\otimes n$} representation.
We show that the magnetic weights of the so-called fundamental $\mathbb{Z}_{2}$
monopoles correspond to the weights of the defining representation
of the dual algebra $so(n)^{\vee}$, and the masses of the nonfundamental
BPS $\mathbb{Z}_{2}$ monopoles are equal to the sum of the masses
of the constituent fundamental monopoles. We also show that the vacua
responsible for the existence of these $\mathbb{Z}_{2}$ monopoles
are present in the Higgs branch of a class of ${\cal N}=2$ $SU(n)$
superconformal field theories. We analyze some dualities these monopoles
may satisfy.

\vfill 


\thispagestyle{empty}
\end{abstract}
\newpage

\section{Introduction}

Electromagnetic duality in Yang-Mills-Higgs theories was initially
proposed by Goddard, Nuyts, and Olive (GNO) \cite{GNO} in gauge theories
with gauge group $G$ spontaneously broken to $G_{0}$ by a scalar
field $\phi$ in an arbitrary representation, in such a way that $\pi_{2}(G/G_{0})$
is nontrivial, which allows the existence of monopole solutions. Soon
after, Montonen and Olive duality was proposed \cite{Monotonen-Olive}
considering a theory with gauge group $SU(2)$ spontaneously broken
to $U(1)$ by a scalar field $\phi$ in the adjoint representation.
Since then, monopole solutions and the electromagnetic duality have
been studied mainly when the scalar field responsible for the symmetry
breaking is in the adjoint representation. In this case, the unbroken
gauge group $G_{0}$ necessarily has a $U(1)$ factor which guarantees
that $\pi_{2}(G/G_{0})=\mathbb{Z}$ and that the theory can have monopole
solutions which we shall call $\mathbb{Z}$ monopoles. On the other
hand, much less is known when $G_{0}$ is semisimple and therefore,
$\phi$ necessarily cannot be in the adjoint representation. In these
cases, a nontrivial $\pi_{2}(G/G_{0})$ will be a cyclic group $\mathbb{Z}_{n}$
or a product of cyclic groups and the monopoles are called $\mathbb{Z}_{n}$
monopoles. Therefore, $\mathbb{Z}_{n}$ monopoles are relevant for
GNO duality when $G_{0}$ is semisimple, which has renewed interest
in the geometric Langlands program \cite{langlands}. These $\mathbb{Z}_{n}$
monopoles were analyzed, for example, in \cite{GNO}\cite{Weinberg}\cite{Bais}
and more recently in \cite{KL2010}. 

One of the main motivations for the study of monopoles and electromagnetic
dualities is their possible application to the problem of confinement
in QCD. Following the ideas of  't Hooft and Mandelstam, the formation
of chromoelectric flux tubes in QCD must be due to a monopole condensate.
However, it is not yet clear if these monopoles are $\mathbb{Z}$
monopoles, $\mathbb{Z}_{n}$ monopoles, or Dirac monopoles. In the
last few years, the ideas of  't Hooft and Mandelstam were applied
to supersymmetric non-Abelian theories  with $\mathbb{Z}$ monopoles.
In particular the confinement of $\mathbb{Z}$ monopoles by the formation
of magnetic flux tubes or $\mathbb{Z}_{n}$ strings in soft broken
${\cal N}=4$ super Yang-Mills theories with arbitrary simple gauge
groups was analyzed in \cite{Kneipp2004}\cite{kneipp2007}and it
was shown that in these theories the tensions of these $\mathbb{Z}_{n}$
strings satisfy the Casimir scaling law in the BPS limit, which is
believed to be the behavior that the chromoelectric flux tubes in
QCD must satisfy\cite{lattice results}. This result indicates that
these $\mathbb{Z}_{n}$ strings can be dual to QCD chromoelectric
strings.

In order to understand the properties of the $\mathbb{Z}_{n}$ monopoles,
in \cite{KL2010} we obtained explicitly the asymptotic form of the
$\mathbb{Z}_{2}$ monopoles in $SU(n)$ Yang-Mills-Higgs theories
with the gauge group broken to $Spin(n)/\mathbb{Z}_{2}$ by a scalar
in the $n\otimes n$ representation of $SU(n)$ or its symmetric part.
In order to obtain these asymptotic forms, we generalized the construction
in \cite{Weinberg} using the fact that the magnetic weights of the
monopoles in this theory must belong to the cosets $\Lambda_{r}(Spin(n)^{\vee})$
or $\lambda_{1}^{\vee}+\Lambda_{r}(Spin(n)^{\vee})$ corresponding
to the two topological sectors associated to the group $\mathbb{Z}_{2}$.
It is important to note that the fact that $\mathbb{Z}_{2}$ monopoles
are associated to $\mathbb{Z}_{2}$ topological sectors does not imply
that they carry nonadditive magnetic charges as we will explain in
sections 2 and 3. We constructed the monopole solutions considering
two symmetry breakings of the algebra $su(n)$ to $so(n)$: one in
which $so(n)$ is invariant under outer automorphism and another in
which it is invariant under Cartan automorphism. In both cases we
associated a $su(2)$ subalgebra, subject to some constraints, to
each weight of the defining representation of the dual algebra $so(n)^{\vee}$
and constructed explicitly the $\mathbb{Z}_{2}$ monopoles called
fundamental monopoles. Using linear combinations of the generators
of these $su(2)$ subalgebras, we were able to construct other $su(2)$
subalgebras and the corresponding $\mathbb{Z}_{2}$ monopoles called
nonfundamental. 

In this paper we write the vacuum solution and the asymptotic forms
for the $\mathbb{Z}_{2}$ monopoles in terms of singlets and triplets
with respect to the corresponding $su(2)$ subalgebras. We calculate
the masses for the BPS monopoles and obtained that the fundamental
BPS $\mathbb{Z}_{2}$ monopoles have the same masses equal to $4\pi v/e$,
where $v$ is the norm of the Higgs vacuum. On the other hand, the
masses of the nonfundamental $\mathbb{Z}_{2}$ monopoles are the sum
of the masses of the constituent fundamental monopoles which is consistent
with the interpretation that the nonfundamental monopoles should be
multimonopoles composed of noninteracting fundamental monopoles, similarly
to what happens for the $\mathbb{Z}$ monopoles \cite{Weinberg fundamental}. 

Exact electromagnetic duality is expected to happen in superconformal
theories (SCFTs), with a vanishing $\beta$ function, like ${\cal N}=4$
super Yang-Mills theories\cite{N=00003D4}, ${\cal N}=2$ $SU(2)$
super Yang-Mills theories with $N_{F}=4$ flavors \cite{Seiberg-Witten2},
etc. More recently, with the works \cite{Argyres-Seiberg}\cite{Gaiotto},
there was some renewed interest with the study of dualities in SCFTs.
The $\mathbb{Z}_{2}$ monopoles cannot exist in ${\cal N}=4$ super
Yang-Mills theories, where all scalars are in the adjoint representation.
Therefore, in order to analyze some possible dualities that $\mathbb{Z}_{2}$
monopoles may satisfy in SCFTs, we consider ${\cal N}=2$ $SU(n)$
super Yang-Mills theories with a hypermultiplet in the $n\otimes n$
representation, which has a vanishing $\beta$ function  and which
we will denote by ${\cal N}=2'$ SCFTs. We showed that its potential
accepts the vacua solutions discussed in the previous sections. These
vacua correspond to certain points of the Higgs branch where the $\mathbb{Z}_{2}$
monopoles can exist. That is different from the Coulomb branch where
the gauge symmetry is generically broken to the maximal torus $U(1)^{r}$
{[}or to $K\times U(1)$ in some specific points{]} and there are
$\mathbb{Z}$ monopoles/dyons everywhere on the Coulomb branch. It
is interesting to note that the BPS equations for the $\mathbb{Z}_{2}$
monopoles do not result on vanishing of any supercharges. Therefore,
even the BPS $\mathbb{Z}_{2}$ monopoles satisfying the first order
BPS equations are in long ${\cal N}=2$ massive supermultiplets, like
the massive gauge fields in this theory. We also showed that this
${\cal N}=2'$ SCFT can have an Abelian Coulomb phase with $\mathbb{Z}_{2}$
monopoles and $\mathbb{Z}$ monopoles. From the results we obtained,
we discussed some possible dualities the $\mathbb{Z}_{2}$ monopoles
may satisfy. 

This paper is organized as follows: we start in Sec. 2 giving a short
review of our generalized construction of the spherically symmetric
$\mathbb{Z}_{n}$ monopole's asymptotic forms. Then, in Sec. 3 obtain
explicitly the asymptotic form of the $\mathbb{Z}_{2}$ monopoles
in $SU(n)$ Yang-Mills-Higgs theories with the gauge group broken
to $Spin(n)/\mathbb{Z}_{2}$ by a scalar in the $n\otimes n$ representation.
We consider two vacua configurations which break $su(n)$ to $so(n)$
where for the first configuration $so(n)$ is invariant under Cartan
automorphism and for the second configuration it is invariant under
outer automorphism. In Sec. 4 we calculate the BPS masses for the
fundamental and nonfundamental $\mathbb{Z}_{2}$ monopoles. In Sec.
5, we show that the vacua responsible for the breaking $SU(n)$ to
$Spin(n)/\mathbb{Z}_{2}$ belong to the Higgs branch of a ${\cal N}=2$
$SU(n)$ SCFT and therefore this theory can have these $\mathbb{Z}_{2}$
monopoles. Finally, in Sec. 6 we discuss some possible dualities these
$\mathbb{Z}_{2}$ monopoles can satisfy.

\section{General properties of $\mathbb{Z}_{n}$ monopoles}

In this section we shall recall some of the principal results of $Z_{n}$
monopoles and fix some conventions. For more details, see \cite{KL2010}.
Let us start by considering a Yang-Mills theory with gauge group $G$
which we shall consider to be simple and simply connected. Let us
also consider that the theory has a scalar field $\phi$ in a representation
$R(G)$ and $\phi_{0}$ is a vacuum configuration which spontaneously
breaks $G$ to $G_{0}$ such that $\pi_{2}(G/G_{0})$ is nontrivial,
and therefore allows the existence of monopoles. Let us denote%
\footnote{We shall adopt the convention of using capital letters to denote Lie
groups and lower letters for  Lie algebras.%
} by $g$ the algebra formed by the generators of $G$ and $g_{0}$
the generators of $G_{0}$. Note that in general, the elements of
the Cartan subalgebra (CSA) of $g_{0}$ do not necessarily belong
to the CSA of $g$. Therefore, we shall denote by $H_{i}$ and \emph{$E_{\alpha}$,}
respectively, the CSA's generators and the step operators of $g$,
and $h_{i}$ and $f_{\alpha}$ the corresponding generators of $g_{0}$.
We shall adopt the convention that in the Cartan-Weyl basis, the commutation
relations read\begin{eqnarray}
\left[H_{i},E_{\alpha}\right] & = & (\alpha)^{i}E_{\alpha},\label{eq:2.1}\\
\left[E_{\alpha},E_{-\alpha}\right] & = & \frac{2\alpha\cdot H}{\alpha^{2}}.\nonumber \end{eqnarray}

We shall denote by $\alpha_{i},\, i=1,\,2,\,...,\, r=\mbox{rank\,}g$,
the simple roots of $g$ and $\lambda_{i},\, i=1,\,2,\,...,\, r$,
the fundamental weights of $g$. Moreover, we shall denote\begin{equation}
\alpha_{i}^{\vee}=\frac{2\alpha_{i}}{\alpha_{i}^{2}},\,\,\,\,\,\,\,\lambda_{i}^{\vee}=\frac{2\lambda_{i}}{\alpha_{i}^{2}},\label{2.2}\end{equation}
as the simple coroots and fundamental coweights, respectively. They
are simple roots and fundamental weights of the dual algebra $g^{\vee}$
and satisfy the relations $\alpha_{i}\cdot\lambda_{j}^{\vee}=\alpha_{i}^{\vee}\cdot\lambda_{j}=\delta_{ij}$.

The asymptotic condition $\mathcal{D}_{i}\phi=0$ for finite energy
configurations implies that asymptotically we can write \begin{equation}
\phi(\theta,\varphi)=g(\theta,\varphi)\phi_{0},\label{2.3}\end{equation}
where $\theta$ and $\varphi$ are the angular spherical coordinates
and $g(\theta,\varphi)\in G$. Then, the asymptotic form of the magnetic
field of the monopoles can be written as \cite{GNO}\begin{equation}
B_{i}(\theta,\varphi)=\frac{x_{i}}{2er^{3}}g(\theta,\varphi)\,\omega\cdot h\, g(\theta,\varphi)^{-1}\label{2.4}\end{equation}
where $\omega$ is a real vector called magnetic weight and $h_{i}$
belongs to the CSA of $g_{0}$. 

Note that when the gauge group $G$ is broken by a scalar field in
the adjoint representation, the unbroken gauge group $G_{0}$ always
has a $U(1)$ factor generated by the scalar field vacuum solution
$\Phi_{0}=\phi_{0a}T_{a}$ and we can define an Abelian magnetic charge
for the monopole associated to this $U(1)$ factor\[
\mathtt{g}=\frac{1}{\left|\Phi_{0}\right|}\ointop_{S_{\infty}^{2}}dS_{i}\mbox{Tr}\left(B_{i}\Phi\right)=\frac{1}{\left|\Phi_{0}\right|}\ointop_{S_{\infty}^{2}}dS_{i}\mbox{Tr}\left(B_{i}g(\theta,\varphi)\Phi_{0}g(\theta,\varphi)^{-1}\right).\]
On the other hand, when $\phi$ is not in the adjoint representation,
we cannot define the above charge, but we can define magnetic charges
associated to the CSA generators $h_{a}$ of the unbroken group $G_{0}$
as%
\footnote{Remember that when we have a monopole solution, the unbroken group
is not fixed but varies with the space direction within $G$ by conjugation
$g(\theta,\varphi)G_{0}g(\theta,\varphi)^{-1}$ \cite{GO review}%
} \begin{equation}
\mathtt{g}_{a}=\ointop_{S_{\infty}^{2}}dS_{i}\mbox{Tr}\left(B_{i}g(\theta,\varphi)h_{a}g(\theta,\varphi)^{-1}\right)=\frac{2\pi}{e}\omega_{a}.\label{eq:2.5}\end{equation}
Therefore, these magnetic charges are proportional to the components
of the magnetic weight associated to a monopole.

Considering that $G_{0}$ is semisimple, it can be written as\[
G_{0}=\widetilde{G_{0}}/K(G_{0})\]
where $\widetilde{G_{0}}$ is the universal covering group of $G_{0}$
and $K(G_{0})$ is the kernel of the homomorphism $\widetilde{G_{0}}\rightarrow G_{0}$.
One can show that $K(G_{0})$ is a discrete subgroup of the center
of $\widetilde{G_{0}}$, which we will call $Z(\widetilde{G_{0}})$.
Therefore, when $G_{0}$ is semisimple, the topological charge sectors
of the theory are associated to \begin{equation}
\pi_{2}(G/G_{0})=\pi_{1}(G_{0})=K(G_{0})\subset Z(\widetilde{G_{0}}).\label{eq:2.6a}\end{equation}
Hence, $\pi_{2}(G/G_{0})$ is a cyclic group $\mathbb{Z}_{n}$, or
a product of cyclic groups, and the monopoles are called $\mathbb{Z}_{n}$
monopoles. 

Now, the center of a group $\widetilde{G_{0}}$ is a discrete group
isomorphic to the classes\begin{eqnarray}
Z(\widetilde{G_{0}}) & = & \left\{ \exp\left[2\pi i\Lambda_{r}(G_{0}^{\vee})\cdot h\right],\,\exp\left[2\pi i\left(\lambda_{\tau(0)}^{\vee}+\Lambda_{r}(G_{0}^{\vee})\right)\cdot h\right],\,\ldots\,,\right.\label{eq:2.6}\\
 &  & \left.\ldots,\,\exp\left[2\pi i\left(\lambda_{\tau^{n}(0)}^{\vee}+\Lambda_{r}(G_{0}^{\vee})\right)\cdot h\right]\right\} ,\nonumber \end{eqnarray}
where $\Lambda_{r}(G_{0}^{\vee})$ is the root lattice of $G_{0}^{\vee}$,
the dual group of $G_{0}$, and the fundamental coweights $\lambda_{\tau^{q}(0)}^{\vee}$
are associated to the nodes of the extended Dynkin diagram of $G_{0}$
related to the node $0$ by a symmetry transformation, as explained
in detail in \cite{olive-turok 1983}. The relation (\ref{eq:2.6})
is due to the fact that the quotient $\Lambda_{w}(\widetilde{G}_{0}^{\vee})/\Lambda_{r}(G_{0}^{\vee})$
can be represented by the cosets \begin{equation}
\Lambda_{r}(G_{0}^{\vee}),\quad\lambda_{\tau(0)}^{\vee}+\Lambda_{r}(G_{0}^{\vee}),\quad\lambda_{\tau^{2}(0)}^{\vee}+\Lambda_{r}(G_{0}^{\vee}),\quad\dots\quad,\lambda_{\tau^{n}(0)}^{\vee}+\Lambda_{r}(G_{0}^{\vee}).\label{eq:2.7}\end{equation}
Since $K(G_{0})\subset Z(\widetilde{G_{0}})$, the topological charge
sectors (\ref{eq:2.6a}) are associated to the elements of (\ref{eq:2.6})
which are in the kernel of the homomorphism $\widetilde{G_{0}}\rightarrow G_{0}$. 

The group element $g(\theta,\varphi)$ must satisfy the relation \cite{GNO}
\begin{equation}
g\left(\pi,0\right)^{-1}g(\pi,2\pi)=\widetilde{\exp}\left[2\pi i\omega\cdot h\right]\in K(G_{0})\subset Z(\widetilde{G_{0}})\label{eq:2.8}\end{equation}
where $\widetilde{\exp}$ denotes the exponential mapping in $\widetilde{G_{0}}$.
Hence, $\widetilde{\exp}\left[2\pi i\omega\cdot h\right]$ must be
in one of the classes of (\ref{eq:2.6}) associated to $K(G_{0})$.
Therefore, the magnetic weights $\omega$ must be only in the cosets
associated to the kernel $K(G_{0})$ and the $\mathbb{Z}_{n}$ monopoles
will be in the same topological sector if their associated magnetic
weights $\omega$ are in the same coset \cite{KL2010}. The coset
$\Lambda_{r}(G_{0}^{\vee})$ corresponds to the trivial element $\mathbbm{1}$
of the group $\mathbb{Z}_{n}$, and monopoles with magnetic weights
in this coset belong to the trivial topological sector. Note that
two $\mathbb{Z}_{n}$ monopoles in the same topological sector, i.e.,
those associated to magnetic weights $\omega^{A}$ and $\omega^{B}$
in the same coset, does not imply that they are the same monopole
since they have different asymptotic magnetic fields (\ref{2.4}).
However, some of these monopole solutions, but not all of them, can
be related by gauge transformations. These gauge transformations have
the effect to produce Weyl reflections on the magnetic weights of
the monopoles \cite{GNO}.

Let us now consider a generator \[
T_{3}^{\beta}=\frac{\beta\cdot h}{2}\]
such that $\beta$ is a vector that belongs to one of the cosets associated
to $K(G_{0})$, that is, $\beta$ can be a magnetic weight. Let us
also consider that exist two other generators $T_{1}^{\beta},T_{2}^{\beta}\notin g_{0}$
exist, which together with $T_{3}^{\beta}$ form a $su(2)$ subalgebra
\[
\left[T_{i}^{\beta},T_{j}^{\beta}\right]=i\epsilon_{ijk}T_{k}^{\beta},\]
which we shall denote $su(2)_{\beta}$. Since $\widetilde{\exp}\left[2\pi i\beta\cdot h\right]\in K(G_{0})$,
then $\widetilde{\exp}\left[2\pi iq\beta\cdot h\right]\in K(G_{0})$
where $q\in\mathbb{Z}$. Therefore, $q\beta$ is also in one of the
cosets associated to $K(G_{0})$. Since we are interested in the study
of fundamental monopoles, we shall consider solutions with spherically
symmetric asymptotic forms. As in \cite{KL2010}, from these generators,
we shall obtain explicit monopole asymptotic forms with spherical
symmetry using a generalization of the construction in \cite{Weinberg},
writing the group element $g(\theta,\varphi)$ as \begin{equation}
g(\theta,\varphi)=\exp(-i\varphi qT_{3}^{\beta})\exp(-i\theta T_{2}^{\beta})\exp(i\varphi qT_{3}^{\beta}),\label{monopolos11}\end{equation}
which satisfies\begin{equation}
g\left(\pi,0\right)^{-1}g(\pi,2\pi)=\widetilde{\exp}\left[2\pi iq\beta\cdot h\right]\in K(G_{0}).\label{eq:monopolos11a}\end{equation}
Therefore, the monopole associated to this group element has magnetic
weight $\omega=q\beta$. Hence, for each integer $q$ and $su(2)_{\beta}$
subalgebra with $T_{3}^{\beta}$ satisfying condition (\ref{eq:2.8}),
we can associate a $\mathbb{Z}_{n}$ monopole. A very important difference
from the construction in \cite{Weinberg} is that in our construction
the monopole topological sectors are associated to the cosets (\ref{eq:2.7})
and not to the integer $q$ and therefore, monopoles associated to
magnetic weights with the same integer $q$ are not necessarily in
the same topological sector. As a consequence, from our generalized
construction we obtain many more solutions. One can think of the monopoles
associated to a $su(2)_{\beta}$ and with $|q|>1$ as superpositions
of $|q|$ monopoles with $|q|=1$ associated to the same $su(2)_{\beta}$.
Similarly to \cite{Weinberg}, we consider that a monopole associated
to a $su(2)_{\beta}$ subalgebra with a negative integer $-q$ is
the antimonopole of the monopole with positive integer $q$ and is
associated to the same $su(2)_{\beta}$. It is interesting to note
that in particular, a $\mathbb{Z}_{2}$ monopole and its antiparticle
are in the same topological sector, but if one has magnetic weight
$q\beta$, the other has $-q\beta$. 

Using the identity, for $i\neq j$,\begin{equation}
\exp(iaT_{j})T_{i}\exp(-iaT_{j})=(\cos a)T_{i}+(\sin a)\epsilon_{ijk}T_{k},\label{eq:2.9}\end{equation}
where $T_{i}$, $i=1,2,3$ form an arbitrary $su(2)$ subalgebra,
we can rewrite the asymptotic form (\ref{2.4}) for the magnetic field
with $\omega=q\beta$ and $g(\theta,\varphi)$ given by (\ref{monopolos11})
as\begin{equation}
B_{i}^{^{(q)}}(\theta,\varphi)=\frac{qx_{i}}{er^{3}}\left[T_{3}^{\beta}\cos\theta+\sin\theta\left(T_{1}^{\beta}\cos q\varphi+T_{2}^{\beta}\sin q\varphi\right)\right].\label{eq:2.11}\end{equation}

One can obtain this asymptotic form from the gauge field configuration
\cite{Weinberg} \begin{equation}
W_{i}(\theta,\varphi)=g(\theta,\varphi)W_{i}^{\mbox{\scriptsize string}}g(\theta,\varphi)^{-1}-\frac{i}{e}(\partial_{i}g(\theta,\varphi))g(\theta,\varphi)^{-1},\label{monopolos13-1}\end{equation}
 with \[
W_{r}^{\scriptsize\mbox{string}}=W_{\theta}^{\mbox{\scriptsize string}}=0,\]
 \[
W_{\varphi}^{\scriptsize\mbox{string}}=-\frac{qT_{3}^{\beta}}{er}\frac{(1-\cos\theta)}{\sin\theta},\]
which gives \begin{subequations}\label{vector}

\begin{eqnarray}
W_{r}(\theta,\varphi) & = & 0,\\
W_{\theta}(\theta,\varphi) & = & \frac{1}{er}\left(T_{1}^{\beta}\sin q\varphi-T_{2}^{\beta}\cos q\varphi\right),\\
W_{\varphi}(\theta,\varphi) & = & \frac{q}{er}\left[-T_{3}^{\beta}\sin\theta+\cos\theta\left(T_{1}^{\beta}\cos q\varphi+T_{2}^{\beta}\sin q\varphi\right)\right].\end{eqnarray}
\end{subequations}

\section{$\mathbb{Z}_{2}$ monopoles in $SU(n)$ Yang-Mills-Higgs theories}

Let us consider a Yang-Mills-Higgs theory with gauge group $SU(n)$
and a scalar field $\phi$ in the direct product representation $n\otimes n$
of $SU(n)$. In order for $\mathbb{Z}_{2}$ monopoles to exist, in
\cite{KL2010} we find vacuum solutions $\phi_{0}$ which break \begin{equation}
SU(n)\rightarrow\frac{Spin(n)}{\mathbb{Z}_{2}}\label{eq:3.0}\end{equation}
for $n\geq3$, where $Spin(n)$ is the covering group of $SO(n)$
and is associated to the algebra $so(n)$. We consider two different
vacua: for one vacuum, the unbroken $so(n)$ is the subalgebra of
$su(n)$ invariant under Cartan automorphism and for the second vacuum,
$so(n)$ is the subalgebra invariant under outer automorphism, and
in this case $n$ must be odd. In both cases, the kernel $K(G_{0})=\mathbb{Z}_{2}$
is associated to the cosets \begin{equation}
\Lambda_{r}(Spin(n)^{\vee}),\quad\lambda_{1}^{\vee}+\Lambda_{r}(Spin(n)^{\vee}),\label{eq:3.1}\end{equation}
where $\lambda_{1}$ is a fundamental weight of the $so(n)$ subalgebra,
using the convention of \cite{KL2010}. The first coset is associated
to the trivial topological sector. As explained in detail in Sec.
6 of \cite{KL2010}, if we consider two $\mathbb{Z}_{2}$ monopoles
with magnetic weights $\omega^{(A)}$ and $\omega^{(B)}$ belonging
to the coset $\lambda_{1}^{\vee}+\Lambda_{r}(Spin(n)^{\vee}),$ and
therefore belonging to the nontrivial topological sector, then the
monopole composed by these two monopoles will have magnetic weight
$\omega^{(A)}+\omega^{(B)}$, which belongs to $\Lambda_{r}(Spin(n)^{\vee})$
{[}since $2\lambda_{1}^{\vee}\in\Lambda_{r}(Spin(n)^{\vee})${]} and
hence to the trivial topological sector. It means that the $\mathbb{Z}_{2}$
monopole carries an additive magnetic charge, since it is proportional
to its magnetic weight, and the $\mathbb{Z}_{2}$ topological charge
of a monopole is related to the exponential of its magnetic weight
by Eq. (\ref{eq:2.6}).

Before we consider these two symmetry breakings, let us obtain some
Lie algebra results for the $n\otimes n$ representation, which will
be useful in the next sections. Let us denote by $\left|e_{l}\right\rangle $,
$l=1,2,\,...,n$, the weight states of the $n$-dimensional representation
of $su(n)$. In this representation, the generators of $su(n)$ can
be written in terms of the $n\times n$ matrices $E_{ij}$ with components
$\left(E_{ij}\right)_{kl}=\delta_{ik}\delta_{jl}$ or \begin{equation}
E_{ij}\left|e_{j}\right\rangle =\left|e_{i}\right\rangle .\label{eq:3.3}\end{equation}
In this case, the basis elements of the CSA of $su(n)$ correspond
to the traceless combinations $E_{ii}-E_{i+1,i+1}$, for $i=1,2,\,...,n-1$.
The generator $E_{ij},\,\, i\neq j$, is the step operator associated
to the root $e_{i}-e_{j}$, where $e_{i}$ is an orthonormal vector
in the $n$-dimensional vector space. 

In the representation $n\otimes n$, the weight states are $\left|e_{i}\right\rangle \otimes\left|e_{j}\right\rangle ,\, i,j=1,2,\,...,n$,
and the generators can be written as \[
\mathbb{E}_{ij}=E_{ij}\otimes1+1\otimes E_{ij}.\]
In this representation, for a root $\beta=e_{i}-e_{j}$ of $su(n)$,
we can associate a $su(2)_{\beta}$ subalgebra\begin{eqnarray}
T_{3}^{\beta} & = & \frac{\beta\cdot H}{2}=\frac{1}{2}\left(\mathbb{E}_{ii}-\mathbb{E}_{jj}\right),\nonumber \\
T_{1}^{\beta} & = & \frac{E_{\beta}+E_{-\beta}}{2}=\frac{1}{2}\left(\mathbb{E}_{ij}+\mathbb{E}_{ji}\right),\label{eq:3.5}\\
T_{2}^{\beta} & = & \frac{E_{\beta}-E_{-\beta}}{2i}=\frac{1}{2i}\left(\mathbb{E}_{ij}-\mathbb{E}_{ji}\right).\nonumber \end{eqnarray}
Adopting the notation $\left|i,j\right\rangle \equiv\left|e_{i}\right\rangle \otimes\left|e_{j}\right\rangle $,
we can define the weight states\begin{eqnarray}
\left|0\right\rangle _{\beta,1} & = & \frac{1}{\sqrt{2}}\left(\left|j,j\right\rangle -\left|i,i\right\rangle \right),\nonumber \\
\left|0\right\rangle _{\beta,2} & = & \frac{i}{\sqrt{2}}\left(\left|j,j\right\rangle +\left|i,i\right\rangle \right),\label{eq:3.51}\\
\left|0\right\rangle _{\beta,3} & = & \frac{1}{\sqrt{2}}\left(\left|i,j\right\rangle +\left|j,i\right\rangle \right),\nonumber \end{eqnarray}
where $\left|0\right\rangle _{\beta,i}$ is eigenvector of $T_{i}^{\beta}$
with vanishing eigenvalue and one can check \begin{equation}
T_{i}^{\beta}|0\rangle_{\beta,j}=i\sum_{k}\epsilon_{ijk}|0\rangle_{\beta,k}.\label{eq:3.52}\end{equation}
Remembering that for an arbitrary Lie algebra, a weight state $|T_{i}\rangle$
of the adjoint representation is associated to a generator $T_{i}$
through the relation \[
T_{i}|T_{j}\rangle=i\sum_{k}f_{ijk}|T_{k}\rangle=|\left[T_{i},T_{j}\right]\rangle,\]
where $f_{ijk}$ are the structure constants of the algebra. Therefore,
from Eq. (\ref{eq:3.52}) we can conclude that the weight states (\ref{eq:3.51})
form an adjoint or triplet representation of the $su(2)_{\beta}$
subalgebra (\ref{eq:3.5}) and we can associate $|0\rangle_{\beta,j}$
to $T_{j}^{\beta}$.

\subsection{$so(2m+1)$ invariant under outer automorphism}

Let us consider first the case where $so(2m+1)$ is the invariant
subalgebra of $su(2m+1)$ under outer automorphism. In this case,
the CSA of $so(2m+1)$ is inside the CSA of $su(2m+1)$, as explained
in detail in \cite{KL2010}. The vacuum configuration which breaks
$su(2m+1)$ to this $so(2m+1)$ subalgebra is \cite{KL2010}\begin{equation}
\phi_{0}=\frac{v}{\sqrt{2}}\sum_{l=1}^{2m+1}\left(-1\right)^{l+1}\left|l,2m+2-l\right\rangle ,\label{eq:3.2}\end{equation}
where $v$ is a real constant.

Let us now analyze the possible $\mathbb{Z}_{2}$ monopole solutions
of the theory. Since for the moment we are interested in the so-called
fundamental monopoles, we shall consider that $q=1$. The monopoles
associated to the nontrivial topological sector must have magnetic
weights $\beta$ in the coset $\lambda_{1}^{\vee}+\Lambda_{r}(Spin(n)^{\vee}).$
This condition is written in terms of coweights and coroots of the
subalgebra $so(2m+1)$. We showed that this condition can be written
in terms of roots of $su(2m+1)$ as \begin{equation}
\beta\in\left[\sum_{i=1}^{m-1}c_{i}\left(\alpha_{i}+\alpha_{2m+1-i}\right)\right]+\left(2c_{m}+1\right)\left(\alpha_{m}+\alpha_{m+1}\right)\label{eq:3.4}\end{equation}
where $c_{i}$ are arbitrary integers and $\alpha_{i}$ are simple
roots of $su(2m+1)$. On the other hand, the monopoles associated
to the trivial topological sector must have magnetic weights $\beta$
in the coset $\Lambda_{r}(Spin(n)^{\vee})$. This condition can be
written in terms of roots of $su(2m+1)$ as\begin{equation}
\beta\in\left[\sum_{i=1}^{m-1}c_{i}\left(\alpha_{i}+\alpha_{2m+1-i}\right)\right]+2c_{m}\left(\alpha_{m}+\alpha_{m+1}\right),\label{eq:3.4b}\end{equation}
with $c_{i}$ being integers. Therefore, $\beta$ can only be in the
particular subspace of $\Lambda_{r}\left(SU(2m+1)\right)$ which is
the union of the subspaces given by conditions (\ref{eq:3.4}) and
(\ref{eq:3.4b}). In order to construct $su(2)_{\beta}$ subalgebras,
we consider that $\beta$ is a root of $su(2m+1)$ in this subspace.
In this case, we can consider a $su(2)_{\beta}$ subalgebra of the
form of (\ref{eq:3.5}) which satisfies all the properties discussed
before. The only roots of $su(2m+1)$ which satisfy condition (\ref{eq:3.4})
of being in the nontrivial sector, are \cite{KL2010}\begin{equation}
\pm\left(\alpha_{p}+\alpha_{p+1}+\ldots+\alpha_{2m+1-p}\right)\,,\, p=1,2,\ldots,m.\label{eq:3.6}\end{equation}
On the other hand, there is no root of $su(2m+1)$ which satisfies
condition (\ref{eq:3.4b}). We constructed other $su(2)_{\beta}$
subalgebras associated to other elements in the cosets (\ref{eq:3.1}).
However, in all the cases we found, the generators were always linear
combinations of the generators of (\ref{eq:3.5}). Therefore, we call
fundamental $\mathbb{Z}_{2}$ monopoles, the monopoles associated
to the $su(2)_{\beta_{p}}$ subalgebras (\ref{eq:3.5}) with $\beta$
being one of the $2m$ roots (\ref{eq:3.6}), similarly to the nomenclature
used in \cite{Weinberg fundamental} for the $\mathbb{Z}$ monopoles.
All of these fundamental monopoles are in the nontrivial topological
sector. These $2m$ roots can be written as the weights of the $2m$-dimensional
defining representation of $so(2m+1)^{\vee}=sp(2m)$. 

Using the fact that the simple roots of $su(2m+1)$ can be written
as $\alpha_{p}=e_{p}-e_{p+1}$, we can write these $2m$ roots, or
magnetic weights, as \[
\beta_{p}=e_{p}-e_{2m+2-p},\]
for $p=1,2,\ldots,m,m+2,m+3,\ldots,2m+1$. We can write the generators
of the $su(2)_{\beta}$ subalgebra (\ref{eq:3.5}) associated to $\beta_{p}$
in the $n\otimes n$ representation as \begin{eqnarray}
T_{3}^{\beta_{p}} & = & \frac{1}{2}(\mathbb{E}_{p,p}-\mathbb{E}_{2m+2-p,2m+2-p}),\nonumber \\
T_{1}^{\beta_{p}} & = & \frac{1}{2}\left(\mathbb{E}_{p,2m+2-p}+\mathbb{E}_{2m+2-p,p}\right),\label{eq:3.8}\\
T_{2}^{\beta_{p}} & = & \frac{1}{2i}\left(\mathbb{E}_{p,2m+2-p}-\mathbb{E}_{2m+2-p,p}\right),\nonumber \end{eqnarray}
and the corresponding weight vectors \begin{eqnarray}
|0\rangle_{p,1} & = & (-1)^{p+1}\frac{1}{\sqrt{2}}\left(|2m+2-p,2m+2-p\rangle-|p,p\rangle\right),\nonumber \\
|0\rangle_{p,2} & = & (-1)^{p+1}\frac{i}{\sqrt{2}}\left(|2m+2-p,2m+2-p\rangle+|p,p\rangle\right),\\
|0\rangle_{p,3} & = & (-1)^{p+1}\frac{1}{\sqrt{2}}(|p,2m+2-p\rangle+|2m+2-p,p\rangle),\nonumber \end{eqnarray}
 which are in the adjoint representation of the $su(2)_{\beta_{p}}$
subalgebra (\ref{eq:3.8}) and satisfy (\ref{eq:3.52}).

We can write the vacuum configuration (\ref{eq:3.2}) as\foreignlanguage{american}{
\begin{equation}
\phi_{0}=|0\rangle_{p,0}+v|0\rangle_{p,3},\label{monopolobps5}\end{equation}
}where \[
|0\rangle_{p,0}=\frac{v}{\sqrt{2}}\sum_{l\neq p,2m+2-p}(-1)^{l+1}|l,2m+2-l\rangle,\]
is a singlet of $su(2)_{\beta_{p}}$. Note that the $n\otimes n$
representation decomposes into a direct sum of irreducible representations
of $su(2)_{\beta_{p}}$, and since $T_{3}^{\beta_{p}}\phi_{0}=0$,
$\phi_{0}$ must be a linear combination of weights with zero eigenvalues,
which necessarily belong to odd-dimension irreducible representations
of $su(2)_{\beta_{p}}$. 

From relation (\ref{eq:3.52}), it follows that, for $i\neq j$, \[
\exp\left(iaT_{i}^{\beta_{p}}\right)|0\rangle_{p,j}=\cos a|0\rangle_{p,j}-\sin a\sum_{k}\epsilon_{ijk}|0\rangle_{p,k},\]
where $a$ is an arbitrary constant. Hence, acting with the group
element (\ref{monopolos11}) on $|0\rangle_{p,3}$ we obtain \begin{eqnarray*}
g(\theta,\varphi)|0\rangle_{p,3} & = & \cos\theta|0\rangle_{p,3}+\sin\theta\left\{ \cos q\varphi|0\rangle_{p,1}+\sin q\varphi|0\rangle_{p,2}\right\} \end{eqnarray*}
 and therefore, the asymptotic form (\ref{2.3}) for the scalar field
of the $\mathbb{Z}_{2}$ monopole can be written as\begin{equation}
\phi_{(q)}(\theta,\varphi)=|0\rangle_{p,0}+v\left\{ \cos\theta|0\rangle_{p,3}+\sin\theta\left[\cos q\varphi|0\rangle_{p,1}+\sin q\varphi|0\rangle_{p,2}\right]\right\} .\label{eq:monopolbps8}\end{equation}
 In particular, for $q=1$ we get \[
\phi(\theta,\varphi)=|0\rangle_{p,0}+v\sum_{a=1}^{3}\frac{x_{a}}{r}|0\rangle_{p,a},\]
which has the same form of hedgehog as the $\mathbb{Z}$ monopoles.

From the above asymptotic form, we can propose for the scalar field
the ansatz \begin{equation}
\phi_{(q)}(r,\theta,\varphi)=\phi_{sing}+\phi_{(q)}(r,\theta,\varphi)_{trip},\label{monopolobps9}\end{equation}
 where \begin{subequations}\label{monopolobps12} \begin{align}
\phi_{sing} & =|0\rangle_{p,0},\\
\phi_{(q)}(r,\theta,\varphi)_{trip} & =\sum_{a=1}^{3}\Phi_{(q)}(r,\theta,\varphi)_{a}|0\rangle_{p,a},\\
\Phi_{(q)}(r,\theta,\varphi)_{a} & =f(r)v\left(\sin\theta\cos q\varphi,\sin\theta\sin q\varphi,\cos\theta\right),\end{align}
 \end{subequations} with $f(r)$ being a real function such that
$f(r\rightarrow\infty)=1$ and $f(r=0)=0$, in order to avoid a singularity
at the origin. 

As usual for the adjoint representation, using the association $|0\rangle_{\beta,j}$
to $T_{j}^{\beta}$, we can define the scalar field taking values
in the algebra \begin{equation}
\Phi_{(q)}(r,\theta,\varphi)=\sum_{a=1}^{3}\Phi_{(q)}(r,\theta,\varphi)_{a}T_{a}^{\beta_{p}}.\label{monopolobps13}\end{equation}

Then, using the property that group elements $R(g)_{ij}$ in the
adjoint representation satisfy \[
T_{i}R(g)_{ij}=gT_{j}g^{-1},\]
and the fact that $\phi_{(q)}(r,\theta,\varphi)_{trip}=vf(r)g(\theta,\varphi)|0\rangle_{p,3}$
, we obtain that \begin{equation}
\Phi_{(q)}(r,\theta,\varphi)=f(r)vg(\theta,\varphi)T_{3}^{\beta_{p}}g(\theta,\varphi)^{-1}.\label{monopolobps10}\end{equation}

We can construct the so-called nonfundamental monopoles from the $su(2)_{\beta}$
subalgebras \cite{KL2010} \begin{eqnarray}
T_{3}^{n_{p}\beta_{p}} & = & \sum_{p}n_{p}T_{3}^{\beta_{p}},\nonumber \\
T_{1}^{n_{p}\beta_{p}} & = & \sum_{p}n_{p}T_{1}^{\beta_{p}},\label{eq:bps20}\\
T_{2}^{n_{p}\beta_{p}} & = & \sum_{p}n_{p}T_{2}^{\beta_{p}},\nonumber \end{eqnarray}
where $n_{p}=0,1$ and the summation is over either to positive or
negative roots. Then, the corresponding triplet vectors are\begin{eqnarray}
|0\rangle_{1} & = & \sum_{p}n_{p}|0\rangle_{p,1},\nonumber \\
|0\rangle_{2} & = & \sum_{p}n_{p}|0\rangle_{p,2},\label{eq:tripleto}\\
|0\rangle_{3} & = & \sum_{p}n_{p}|0\rangle_{p,3},\nonumber \end{eqnarray}
and the singlet is\begin{equation}
|0\rangle_{0}=\phi_{0}+v\sum_{p}n_{p}|0\rangle_{p,3}.\label{eq:singleto}\end{equation}
For nonfundamental monopoles associated to these $su(2)_{\beta}$
subalgebras, we arrive in the same field configurations (\ref{monopolobps9}).

It is interesting to note that, since $\beta_{p}\cdot\beta_{q}=2\delta_{p,q}$\cite{KL2010},
the generators $T_{i}^{\beta_{p}},\, p=1,\,2,\,...,\, m\,\mbox{or}\, m+2,\,...,\,2m+1;\, i=1,\,2,\,3$
satisfy\begin{equation}
\mbox{Tr}\left(T_{i}^{\beta_{p}}T_{j}^{\beta_{q}}\right)=\frac{\lambda}{2}\delta_{ij}\delta_{pq},\label{eq:ortogonalidade}\end{equation}
 where $\lambda$ is the Dynkin index of the representation. Therefore,
the generators $T_{i}^{\beta_{p}}$ form a subset of the basis of
orthogonal generators of $su(2m+1)$.

\subsection{$so(n)$ invariant under Cartan automorphism}

Let us now consider the case where the algebra $su(n)$ is broken
to the subalgebra $so(n)$ which is invariant under Cartan automorphism.
In this case, the vacuum configuration which produces this symmetry
breaking is \begin{equation}
\phi_{0}=\frac{v}{\sqrt{2}}\sum_{k=1}^{n}\left|k,k\right\rangle .\label{eq:3.21}\end{equation}
We must consider the cases $n=2m$ and $n=2m+1$ separately. A basis
for the CSA of these subalgebras $so(2m)$ or $so(2m+1)$, which have
rank $m$, is given by the generators \begin{equation}
h_{k}=-i\left(E_{\alpha_{2k-1}}-E_{-\alpha_{2k-1}}\right),\,\,\, k=1,2,\,\ldots,m,\label{eq:3.21aaa}\end{equation}
where $E_{\alpha_{k}}$ are generators of $su(n)$. The asymptotic
forms for the fundamental $\mathbb{Z}_{2}$ monopoles were constructed
in Ref. \cite{KL2010}. Their magnetic weights $\beta_{p}$ are the
weights of the defining representation of the dual algebra $so(n)^{\vee}$.
We know that $so(2m)^{\vee}=so(2m)$ and $so(2m+1)^{\vee}=sp(2m)$,
and the $2m$ weights of the defining representation of $so(2m)$
and $sp(2m)$ can be written in the basis of orthonormal vectors as
\[
\pm\beta_{p}=\pm e_{p},\,\, p=1,\,2,\,\ldots,\, m.\]
For each weight, we can associate a $su(2)_{\beta_{p}}$ subalgebra
\footnote{For these generators we changed sign conventions with respect to \cite{KL2010}.%
} \begin{eqnarray}
T_{3}^{\pm\beta_{p}} & =\pm & \frac{1}{2}e_{p}\cdot h=\pm\frac{1}{2}h_{p}=\frac{1}{2i}\left(E_{\pm\alpha_{2p-1}}-E_{\mp\alpha_{2p-1}}\right),\nonumber \\
T_{1}^{\pm\beta_{p}} & = & \pm\frac{1}{2}\alpha_{2p-1}\cdot H,\label{eq:3.21a}\\
T_{2}^{\pm\beta_{p}} & = & \frac{1}{2}\left(E_{\pm\alpha_{2p-1}}+E_{\mp\alpha_{2p-1}}\right),\nonumber \end{eqnarray}
where $\alpha_{i}$ is a simple root of $su(n)$ for $n=2m,\,2m+1.$

In order to construct the ansatz, we proceed in the same way as in
the previous case. In the $n\otimes n$ representation, these generators
can be written as\begin{eqnarray*}
T_{3}^{\pm\beta_{p}} & = & \pm\frac{1}{2i}\left(\mathbb{E}_{2p-1,2p}-\mathbb{E}_{2p,2p-1}\right),\\
T_{1}^{\pm\beta_{p}} & = & \pm\frac{1}{2}\left(\mathbb{E}_{2p-1,2p-1}-\mathbb{E}_{2p,2p}\right),\\
T_{2}^{\pm\beta_{p}} & = & \frac{1}{2}\left(\mathbb{E}_{2p-1,2p}+\mathbb{E}_{2p,2p-1}\right).\end{eqnarray*}
The corresponding triplet or adjoint representation weight states
are\begin{eqnarray}
|0\rangle_{p,1} & = & \frac{-i}{\sqrt{2}}\left(|2p-1,2p\rangle+|2p,2p-1\rangle\right),\nonumber \\
|0\rangle_{p,2} & = & \frac{\mp i}{\sqrt{2}}\left(|2p,2p\rangle-|2p-1,2p-1\rangle\right),\label{eq:3.23}\\
|0\rangle_{p,3} & = & \frac{1}{\sqrt{2}}(|2p-1,2p-1\rangle+|2p,2p\rangle).\nonumber \end{eqnarray}
Then, the vacuum configuration (\ref{eq:3.21}) can be written as\foreignlanguage{american}{
\[
\phi_{0}=|0\rangle_{p,0}+v|0\rangle_{p,3},\]
}where \[
|0\rangle_{p,0}=\frac{v}{\sqrt{2}}\sum_{k\neq2p-1,2p}\left|k,k\right\rangle ,\]
is a singlet of $su(2)_{\beta_{p}}$. In order to obtain the fundamental
monopoles associated to these $su(2)_{\beta_{p}}$ subalgebras, we
can repeat all the steps of the previous subsection and arrive in
the monopole ansatz (\ref{monopolobps9}) or (\ref{monopolobps13})
with $T_{i}^{\beta_{p}}$ given by (\ref{eq:3.21a}) .

Similarly to the previous case, we can associate nonfundamental monopoles
to the $su(2)_{\beta}$ subalgebras \cite{KL2010}\begin{eqnarray}
T_{3}^{\pm n_{p}\beta_{p}} & = & \sum_{p}n_{p}T_{3}^{\pm\beta_{p}},\nonumber \\
T_{1}^{\pm n_{p}\beta_{p}} & = & \sum_{p}n_{p}T_{1}^{\pm\beta_{p}},\label{eq:bps21}\\
T_{2}^{\pm n_{p}\beta_{p}} & = & \sum_{p}n_{p}T_{2}^{\pm\beta_{p}}.\nonumber \end{eqnarray}
where $n_{p}=0,1$, and the summation is over either to positive or
negative roots. Then, the corresponding triplet and singlet vectors
have the same form of Eqs.(\ref{eq:tripleto-1}) and (\ref{eq:singleto})
with $|0\rangle_{p,i}$ given by (\ref{eq:3.23}).

Like in the previous case, the generators $T_{i}^{\beta_{p}},\, p=1,\,2,\,...,\, m\,;\, i=1,\,2,\,3$
satisfy the orthogonality condition (\ref{eq:ortogonalidade}) and
therefore they form a subset of the basis of orthogonal generators
of $su(n)$ with $n=2m\,\mbox{or}\,2m+1$ .

Therefore, for both symmetry breakings we can conclude that for any
of the previous $su(2)_{\beta}$ subalgebras and arbitrary integer
$q$ satisfying (\ref{eq:monopolos11a}), $\phi_{(q)}(r,\theta,\varphi)$
can be written as a sum of a singlet (which is constant) and a triplet
of this $su(2)_{\beta}$. Moreover, we can associated to this triplet,
a scalar field taking values in the $su(2)_{\beta}$ \begin{equation}
\Phi_{(q)}(r,\theta,\varphi)=f(r)vg(\theta,\varphi)T_{3}^{\beta_{p}}g(\theta,\varphi)^{-1},\label{eq:tripleto-1}\end{equation}
for the fundamental monopoles and\begin{equation}
\Phi_{(q)}(r,\theta,\varphi)=f(r)vg(\theta,\varphi)T_{3}^{\beta}g(\theta,\varphi)^{-1},\label{eq:tripleto-2}\end{equation}
for the nonfundamental ones, where $f(r\rightarrow\infty)=1$ and
$f(r=0)=0$.

We have seen in both symmetry breakings that for each $su(2)_{\beta}$
subalgebra with generators $T_{i}^{\beta}$ we can use to construct
monopole solutions, there is another $su(2)_{-\beta}$ subalgebra
with generators $T_{i}^{-\beta}$. Hence, we can construct monopoles
with the same magnetic weight $\omega$ considering either $q>0$
and $\beta<0$ or $q<0$ and $\beta>0$. For instance, for algebra
$SU(2)_{-\beta}$ with generators $T_{i}^{-\beta}$ and $q=+1$, we
have\begin{eqnarray*}
B_{i}^{(+)}(\theta,\varphi) & = & \frac{x_{i}}{er^{3}}\left[-T_{3}^{\beta}\cos\theta+\sin\theta\left(T_{1}^{\beta}\cos\varphi+T_{2}^{\beta}\sin\varphi\right)\right],\\
\Phi_{(+)}(\theta,\varphi) & = & v\left[-T_{3}^{\beta}\cos\theta+\sin\theta\left(T_{1}^{\beta}\cos\varphi+T_{2}^{\beta}\sin\varphi\right)\right],\end{eqnarray*}
while for algebra $SU(2)_{\beta}$ with generators $T_{i}^{\beta}$
and $q=-1$ we get, for the case that $so(n)$ is invariant under
outer automorphism,

\begin{eqnarray*}
B_{i}^{(-)}(\theta,\varphi) & = & \frac{x_{i}}{er^{3}}\left[-T_{3}^{\beta}\cos\theta+\sin\theta\left(-T_{1}^{\beta}\cos\varphi+T_{2}^{\beta}\sin\varphi\right)\right],\\
\Phi_{(-)}(\theta,\varphi) & = & v\left[T_{3}^{\beta}\cos\theta+\sin\theta\left(T_{1}^{\beta}\cos\varphi-T_{2}^{\beta}\sin\varphi\right)\right],\end{eqnarray*}
where we have used that $T_{3}^{-\beta}=-T_{3}^{\beta}$, $T_{1}^{-\beta}=T_{1}^{\beta}$
and $T_{2}^{-\beta}=-T_{2}^{\beta}$. This suggests the solution obtained
from the subalgebra $su(2)_{-\beta}$ is not the antimonopole for
the one obtained from $su(2)_{\beta}$. The same result is valid for
$so(n)$ invariant under Cartan automorphism, but in these cases $T_{3}^{-\beta}=-T_{3}^{\beta}$,
$T_{1}^{-\beta}=-T_{1}^{\beta}$, and $T_{2}^{-\beta}=-T_{2}^{\beta}$
and the asymptotic fields $B_{i}^{(-)}$ and $\Phi_{(-)}$ will have
a different form. Therefore, the set of fundamental $\mathbb{Z}_{2}$
monopoles with $q=1,$ together with the set of fundamental monopoles
with $q=-1$, has a behavior analogous to particles in two complex
conjugated representations, like the $N$ and $\bar{N}$ of $SU(N)$,
where the antiparticle of a particle in a multiplet is not in the
same multiplet but in the complex conjugate representation.

\section{BPS $\mathbb{Z}_{2}$ monopoles}

\label{sec:limite.bps}

Let us analyze now the BPS $\mathbb{Z}_{2}$ monopoles for this theory.%
\footnote{For simplicity let us abolish the subscript $(q)$ for both the scalar
and the magnetic field.%
} In \cite{Bais}, is given a general procedure to obtain the BPS bound
for $\mathbb{Z}_{n}$ monopoles. However, let us consider a similar
but different procedure for our ansatz. For both symmetry breakings,
the gauge field takes values in a subalgebra $su(2)_{\beta}$, and
$\phi(r,\theta,\varphi)$ can be written as a sum of a singlet (which
is constant) and a triplet of this $su(2)_{\beta}$ subalgebra. Hence,
the action of the covariant derivative on the scalar field gives \[
\mathcal{D}_{\mu}\phi=\mathcal{D}_{\mu}\phi_{sing}+\mathcal{D}_{\mu}\phi_{trip}=\mathcal{D}_{\mu}\phi_{trip},\]
 and then, \[
(\mathcal{D}_{i}\phi)^{\dagger}(\mathcal{D}_{i}\phi)=(\mathcal{D}_{i}\phi_{trip})^{\dagger}(\mathcal{D}_{i}\phi_{trip})=\sum_{a=1}^{3}(\mathcal{D}_{i}\Phi)_{a}(\mathcal{D}_{i}\Phi)_{a},\]
where in the last equality we used the fact that in our ansatz $\Phi_{a}$
is real. Therefore, we obtain that the mass of a static $\mathbb{Z}_{2}$
monopole associated to a magnetic weight $\beta_{p}$ and arbitrary
$q$, for any of the two symmetry breakings, satisfies \begin{eqnarray}
M_{\beta_{p}} & = & \int\left\{ \frac{1}{2}\left[(B_{ia})^{2}+(\mathcal{D}_{i}\phi)^{\dagger}(\mathcal{D}_{i}\phi)\right]+V\right\} d^{3}x,\nonumber \\
 & = & \int\left\{ \frac{1}{2}\sum_{a=1}^{3}\left[(B_{ia})^{2}+(\mathcal{D}_{i}\Phi)_{a}(\mathcal{D}_{i}\Phi)_{a}\right]+V\right\} d^{3}x,\nonumber \\
 & \geq & \pm\int_{S_{\infty}^{2}}\sum_{a=1}^{3}\left(B_{ia}\Phi_{a}\right)d^{2}S_{i},\nonumber \\
 & =\pm & \int_{S_{\infty}^{2}}v\frac{qx_{i}}{er^{3}}d^{2}s_{i}=\pm\frac{4\pi}{e}qv=\frac{4\pi}{e}|q|v,\label{eq:bps1}\end{eqnarray}
where we used the plus or minus sign depending on whether $q$ is
positive or negative, respectively, since the integral in the first
line is greater than zero and in the last line we used the field configurations
\eqref{eq:2.11} and \eqref{monopolobps12}. Notice that we can obtain
the same result using that \[
\sum_{a=1}^{3}\left(B_{ia}\Phi_{a}\right)=Tr\left(B_{i}\Phi\right)=v\frac{qx_{i}}{er^{3}}\]
where the above trace is in the triplet of $su(2)_{\beta_{p}}$ subalgebra
and using Eqs. \eqref{2.4} and \eqref{eq:tripleto-1}. In this case
the group elements $g(\theta,\varphi)$ cancel and therefore, this
bound is valid for more general configurations constructed using group
elements $g(\theta,\varphi)$ other than (\ref{monopolos11}), satisfying
\[
g\left(\pi,0\right)^{-1}g(\pi,2\pi)=\widetilde{\exp}\left[4\pi iT_{3}^{\beta_{p}}\right].\]

Therefore, the masses of the fundamental BPS $\mathbb{Z}_{2}$ monopole
with $q=\pm1$ are \begin{equation}
M_{\beta_{p}}=\frac{4\pi v}{e},\label{bps2}\end{equation}
and they satisfy \begin{subequations}\label{bps3} \begin{align}
E_{a}^{i} & =0,\\
(\mathcal{D}^{0}\Phi)_{a} & =0,\label{eq:BPS equations}\\
B_{a}^{i} & =\pm(\mathcal{D}^{i}\Phi)_{a},\\
V(\phi) & =0,\end{align}
 \end{subequations} where $a=1,2,3$ are associated to the three
generators of the $su(2)_{\beta_{p}}$ subalgebra and the fields associated
to the other generators vanish. From the expression \eqref{bps2},
we can see that it does not depend on $\beta_{p}$, and therefore
all fundamental BPS $\mathbb{Z}_{2}$ monopoles have the same mass.
Since the fields take values only in the $su(2)_{\beta}$ subalgebra
and these BPS equations are the same as in the theory with gauge group
$SU(2)$, it is easy to check that these equations are consistent
with the equations of motion. Moreover, from these equations, as for
the BPS 't Hooft-Polyakov monopole\cite{Bogomolnyi}, we obtain that\begin{eqnarray*}
f(r) & = & \coth\rho-\frac{1}{\rho},\\
a(r) & = & 1-\frac{\rho}{\sinh\rho},\end{eqnarray*}
where $\rho=erv$. 

For the nonfundamental monopoles associated to the $su(2)_{\beta}$
subalgebra (\ref{eq:bps20}) or (\ref{eq:bps21}), the asymptotic
forms of the fields are\begin{eqnarray*}
\Phi_{a} & = & \sum_{p}n_{p}\Phi_{a}^{\beta_{p}},\\
B_{ia} & = & q\sum_{p}n_{p}B_{ia}^{\beta_{p}},\end{eqnarray*}
where $B_{ia}^{\beta_{p}}$ and $\Phi_{a}^{\beta_{p}}$ are asymptotic
forms of the fields of the fundamental monopoles associated to the
$su(2)_{\beta_{p}}$ subalgebras with $q=1$. Therefore, remembering
that $n_{p}=0,1$, we obtain that the BPS limit for the nonfundamental
monopoles is \begin{equation}
M=\frac{4\pi}{e}|q|v\sum_{p}n_{p},\label{bps6}\end{equation}
which is consistent with the interpretation that the nonfundamental
monopoles should be multimonopoles composed of noninteracting fundamental
monopoles, similarly to what happens for the $\mathbb{Z}$ monopoles
\cite{Weinberg fundamental}.

\section{$\mathbb{Z}_{2}$ monopoles at the Higgs branch of ${\cal N}=2'$
SCFTs}

As is well known, there exist some supersymmetric theories with $\mathbb{Z}$
monopoles and a vanishing $\beta$ function, like ${\cal N}=4$ super
Yang-Mills theories, where exact electromagnetic duality is expected
to be valid. Let us now analyze a supersymmetric theory with a vanishing
$\beta$ function and $\mathbb{Z}_{2}$ monopoles. Let us then consider
${\cal N}=2$ $SU(n)$ super Yang-Mills theories with a hypermultiplet
in the $n\otimes n$ representation, which we will call ${\cal N}=2'$
SCFTs. For ${\cal N}=2$ super Yang-Mills, the perturbative $\beta$
function is \begin{equation}
\beta(e)=\frac{2e^{3}}{(4\pi)^{2}}\left(\sum_{i}x_{i}-h^{\vee}\right),\label{super5-1-1}\end{equation}
where $x_{i}$ is the Dynkin index of the hypermultiplets' representations
and $h^{\vee}$ is the dual Coxeter number of the gauge group. For
$G=SU(n)$, $h^{\vee}=n$ and $x(n\otimes n)=n$ since for a representation
$R_{1}\otimes R_{2}$ \[
x(R_{1}\otimes R_{2})=d(R_{1})x(R_{2})+d(R_{2})x(R_{1}),\]
where $d(R)$ is the dimension of the representation $R$ and for
the representation $n$ , $x=1/2$. Therefore, ${\cal N}=2'$ SCFTs
have $\beta(e)=0$. We shall show that its potential accepts the vacua
solutions discussed in the previous sections and therefore $\mathbb{Z}_{2}$
monopoles can exist. The action of the bosonic sector of the $\mathcal{N}=2$
sector of super Yang-Mills with a hypermultiplet can be written as
\begin{equation}
S=\int\left[-\frac{1}{4}G_{\mu\nu}^{a}G^{a\mu\nu}+\frac{1}{2}(\mathcal{D}_{\mu}S)_{a}^{\ast}(\mathcal{D}^{\mu}S)_{a}+\frac{1}{2}(\mathcal{D}_{\mu}\phi_{\alpha})^{\dagger}(\mathcal{D}^{\mu}\phi_{\alpha})+V(S,\phi)\right]d^{4}x,\label{super1}\end{equation}
where $S$ is a scalar field in the adjoint representation, $\phi_{\alpha}$,
$\alpha=1,2$, are complex scalar fields in an arbitrary representation,
and $\sigma^{p}$, $p=1,2,3$, are Pauli matrices. The potential can
be written as \cite{Sohnius} \begin{equation}
V=\frac{e^{2}}{8}\left[(S_{b}^{\ast}if_{abc}S_{c})^{2}+(\phi_{\alpha}^{\dagger}\sigma_{\alpha\beta}^{p}T_{a}\phi_{\beta})^{2}+\frac{4\mu^{2}}{e^{2}}\phi_{\alpha}^{\dagger}\phi_{\alpha}-\frac{4\mu}{e}\phi_{\alpha}^{\dagger}(S+S^{\dagger})\phi_{\alpha}+2\phi_{\alpha}^{\dagger}\left\{ S^{\dagger},S\right\} \phi_{\alpha}\right],\label{super2}\end{equation}
where $\mu$ is a mass parameter which we will show must vanish in
order for the configuration given by Eq. (\ref{eq:3.2}) to be a vacuum
of the theory. We can rewrite the potential as \begin{equation}
V=\frac{1}{2}\left[(d_{a}^{1})^{2}+(d_{a}^{2})^{2}+(D_{a})^{2}+F_{\alpha}^{\dagger}F_{\alpha}\right],\label{super3}\end{equation}
 where \begin{subequations}\label{super4} \begin{align}
d_{a}^{p} & =\frac{e}{2}(\phi_{\alpha}^{\dagger}\sigma_{\alpha\beta}^{p}T_{a}\phi_{\beta}),\quad p=1,2,3,\label{super4a}\\
D_{a} & =\frac{e}{2}(S_{b}^{\ast}if_{abc}S_{c})+d_{a}^{3},\label{super4b}\\
F_{1} & =e\left(S^{\dagger}-\frac{\mu}{e}\right)\phi_{1},\label{super4c}\\
F_{2} & =e\left(S-\frac{\mu}{e}\right)\phi_{2}.\label{super4d}\end{align}

\end{subequations}

\subsection{Non-Abelian Coulomb phase}

\label{sec:coulomb}

In order to produce the gauge symmetry breaking $SU(n)\rightarrow Spin(n)/\mathbb{Z}_{2}$,
which corresponds to the so-called non-Abelian Coulomb phase, we shall
consider the configuration\begin{eqnarray}
\phi_{1vac} & = & \phi_{0},\nonumber \\
\phi_{2vac} & = & 0,\label{eq:coulomb1}\\
S_{vac} & = & 0,\nonumber \end{eqnarray}
where $\phi_{0}$ is one of the two vacua solutions analyzed in the
previous sections. These vacua are in the Higgs branch which does
not receive quantum corrections and the beta function does not receive
nonperturbative corrections \cite{Argyres-Plesser-Seiberg}. We shall
first consider the case of symmetry breaking $su(2m+1)\rightarrow so(2m+1),$
where $so(2m+1)$ is invariant under outer automorphism. Therefore,
we consider that $\phi_{0}$ is given by (\ref{eq:3.2}). Since we
shall consider solutions with $\phi_{2}=0$, we shall use $\phi$
and $F$ to denote $\phi_{1}$ and $F_{1}$, respectively. From Eq.\eqref{super3}
we conclude that in order to obtain $V=0$, we must have that \begin{subequations}\label{coulomb2}
\begin{align}
D & =\frac{e}{2}\left[\sum_{a}(\phi^{\dagger}T_{a}\phi)T_{a}+[S,S^{\dagger}]\right]=0,\label{coulomb2a}\\
F & =e\left(S^{\dagger}-\frac{\mu}{e}\right)\phi=0.\label{coulomb2b}\end{align}
 \end{subequations} We can write \begin{equation}
\sum_{a}(\phi^{\dagger}T_{a}\phi)T_{a}=\sum_{i}(\phi^{\dagger}H_{i}\phi)H_{i}+\frac{1}{2}\sum_{\alpha>0}\alpha^{2}(\phi^{\dagger}E_{\alpha}\phi)E_{-\alpha}.\label{coulomb3}\end{equation}
 Using the fact that the weight state $\left|e_{p}\right\rangle $
of the $(2m+1)$-dimensional representation of $su(2m+1)$ has weight
\[
\omega_{p}\equiv e_{p}-\frac{1}{2m+1}\sum_{i=1}^{2m+1}e_{i}\]
 and $\left\langle e_{p}\right|\left.e_{q}\right\rangle =\delta_{pq}$,
we obtain that \[
\sum_{i}\langle l,2m+2-l|H_{i}\left|p,2m+2-p\right\rangle H_{i}=(\omega_{l}+\omega_{2m+2-l})\cdot H\delta_{lp},\]
 and therefore\begin{eqnarray*}
\sum_{i}(\phi_{vac}^{\dagger}H_{i}\phi_{vac})H_{i} & = & |v|^{2}\sum_{l,p=1}^{2m+1}(-1)^{l+1}(-1)^{p+1}\sum_{i}\langle l,2m+2-l|H_{i}|p,2m+2-p\rangle H_{i},\\
 & = & |v|^{2}\sum_{l=1}^{2m+1}2\omega_{l}\cdot H=0,\end{eqnarray*}
where we used the fact that $\sum_{l}\omega_{l}=0$ for the $(2m+1)$-dimensional
representation of the algebra $su(2m+1)$. On the other hand, for
the step operators we have:\begin{eqnarray*}
\left\langle l,2m+2-l\right|E_{\alpha}\left|p,2m+2-p\right\rangle E_{\alpha} & = & \left(\left\langle e_{l}\right|E_{\alpha}|e_{p}\rangle\langle e_{2m+2-l}|e_{2m+2-p}\rangle+\right.\\
 &  & +\left.\langle e_{l}|e_{p}\rangle\langle e_{2m+2-l}|E_{\alpha}|e_{2m+2-p}\rangle\right)E_{\alpha}=0\end{eqnarray*}
 and therefore \begin{equation}
\sum_{\alpha>0}\alpha^{2}(\phi_{vac}^{\dagger}E_{\alpha}\phi_{vac})E_{-\alpha}=0.\label{coulomb5}\end{equation}
Therefore, we can conclude that the above configuration satisfies
Eq.\eqref{coulomb2a}. It also satisfies \eqref{coulomb2b} if we
consider $\mu=0$ and therefore it is a vacuum of the theory. 

For the case of the breaking of $su(n)$ to $so(n)$ invariant under
Cartan automorphism, one can perform similar calculations using $\phi_{0}$
given by (\ref{eq:3.21}) and verify that it is also a vacuum of this
theory. Hence, the $\mathbb{Z}_{2}$ monopoles analyzed in the previous
sections can exist in this phase of this theory. Note that there will
be $\mathbb{Z}_{2}$ monopoles associated to certain points on the
Higgs branch, differently from the $\mathbb{Z}$ monopoles of the
Coulomb branch where the gauge symmetry is generically broken to the
maximal torus $U(1)^{r}$(or to $K\times U(1)$ in some specific points)
and there are $\mathbb{Z}$ monopoles/dyons everywhere on the Coulomb
branch.

It is interesting to note that, from the ${\cal N}=2$ supersymmetric
variation of the spinoral fields, it is easy to see that the BPS equations
for the $\mathbb{Z}_{2}$ monopoles do not result on vanishing of
any supercharges. This result indicates that, even satisfying the
first order BPS equations (\ref{bps3}), the BPS $\mathbb{Z}_{2}$
monopoles are in a long ${\cal N}=2$ massive supermultiplet and in
principle their masses can receive quantum corrections. It is good
to remember that in this phase where the gauge symmetry breaking is
produced by a scalar which is not in the vector supermultiplet, the
gauge fields which become massive will also belong to a long supermultiplet.
The reason is that in this phase the scalar field is in the hypermultiplet
and is {}``absorbed'' by the gauge fields via Higgs mechanism in
order to form a massive supermultiplet. Therefore, the ${\cal N}=2$
massive vector supermultiplet will be the combination of a massless
vector supermultiplet with a massless hypermultiplet. Note also that
in this phase, the electric and magnetic charges, \begin{eqnarray*}
q & = & \frac{1}{\left|S_{vac}\right|}\ointop_{S_{\infty}^{2}}dS_{i}^{2}G_{a}^{0i}\mbox{Re}\left(S_{a}\right),\\
g & = & \frac{1}{\left|S_{vac}\right|}\ointop_{S_{\infty}^{2}}dS_{i}^{2}\widetilde{G}_{a}^{0i}\mbox{Re}\left(S_{a}\right),\end{eqnarray*}
which appear as central charges, vanish since the scalar field $S$
in the adjoint representation vanishes asymptotically in this phase.

\subsection{Abelian Coulomb phase}

Let us now show that in this theory we can also have the symmetry
breaking sequence\[
SU(2m+1)\rightarrow\frac{Spin(2m+1)}{\mathbb{Z}_{2}}\rightarrow U(1)^{m},\]
which is the Abelian Coulomb phase. In this phase, $\mathbb{Z}_{2}$
monopoles and $\mathbb{Z}$ monopoles as discussed in \cite{bais phase transition,Weinberg}.
Note that this theory can also have an alternative symmetry breaking
(Higgs phase) with confinement of $\mathbb{Z}$ monopoles by $\mathbb{Z}_{2}$
strings \cite{kneipp2001,kneipp2003}. Differently from the non-Abelian
Coulomb phase, in this phase the vacuum moduli space can receive quantum
corrections. For this symmetry breaking, we shall only consider the
case where the subalgebra $so(2m+1)$ is invariant under outer automorphism.
Therefore, we shall consider that $\phi_{1vac}=\phi_{0}$ is given
by (\ref{eq:3.2}) and $\phi_{2vac}=0$. We shall also consider that
$S_{vac}=u\cdot H$ with \begin{equation}
u=a\delta=a\sum_{i=1}^{2m}\lambda_{i}^{\vee},\label{eq:abelian1}\end{equation}
where $a$ is a nonvanishing real constant and $\delta$ is the Weyl
vector of $su(2m+1)$. 

Since $\left[S_{vac},S_{vac}^{\dagger}\right]=0$, it implies that
this new configuration also satisfies Eq. \eqref{coulomb2a}. We now
must show that this configuration is also a solution of \eqref{coulomb2b}.
Substituting (\ref{eq:coulomb1}) in (\ref{coulomb2b}) we obtain\[
F=ve\left\{ \sum_{l=1}^{2m+1}\left(-1\right)^{l+1}\left[u\cdot\left(\omega_{l}+\omega_{2m+2-l}\right)-\frac{\mu}{e}\right]\left|l,2m+2-l\right\rangle \right\} .\]
In order to obtain $F=0$, we must have that\[
u\cdot\left(\omega_{l}+\omega_{2m+2-l}\right)=\frac{\mu}{e}\,\,\,\mbox{for}\,\, l=1,2,\ldots2m+1.\]
It is easy to show that, for $u$ given by \eqref{eq:abelian1}, the
lhs of this equation vanishes for any $l$. Therefore, if we once
more take $\mu=0$, \eqref{eq:coulomb1} with $u$ given by \eqref{eq:abelian1}
is a vacuum solution.

\section{Discussion on duality conjectures}

In order to establish the possible dualities these $\mathbb{Z}_{2}$
monopoles may satisfy, one must determine the gauge multiplet the
fundamental monopoles fill, by doing for example semiclassical quantization.
For this theory that procedure is not simple, since the unbroken gauge
group is non-Abelian, which results in nonnormalizable zero modes
\cite{Weinberg fundamental} and we will leave it for a future work.
Let us therefore discuss some possible dualities the BPS $\mathbb{Z}_{2}$
monopoles may satisfy based on the results we obtained so far. The
particles dual to the BPS $\mathbb{Z}_{2}$ monopoles must be in a
representation with the same weights as the magnetic weights of the
$\mathbb{Z}_{2}$ monopoles. We have seen that, at the classical level,
for a breaking $SU(n)\rightarrow Spin(n)/\mathbb{Z}_{2}$, each fundamental
$\mathbb{Z}_{2}$ monopole is associated to a weight of the defining
representation of $so(n)^{\vee}$ and all of them have the same classical
mass $M_{\beta_{p}}=4\pi v/e$ in the BPS case. Let us consider for
simplicity the even case, $n=2m$, where $so(2m)$ must be an invariant
subalgebra of $su(2m)$ under Cartan involution. In this case $so(2m)^{\vee}=so(2m)$
and the defining representation has dimension $2m$. Therefore, there
are $2m$ fundamental $\mathbb{Z}_{2}$ monopoles. Let us consider
that the dual theory has the same symmetry breaking pattern \begin{equation}
SU(2m)\rightarrow\frac{Spin(2m)}{\mathbb{Z}_{2}},\label{eq:6.2}\end{equation}
with the same vacuum (\ref{eq:3.21}) and also with $\mathbb{Z}_{2}$
monopoles. For this symmetry breaking we have the branchings \begin{eqnarray*}
su(2m) & \rightarrow & so(2m)\\
(1,0,\ldots,0,0)_{su} & \rightarrow & (1,0,\ldots,0)_{so}\,\,,\\
(1,0,\ldots,0,1)_{su} & \rightarrow & (0,1,0,...,0)_{so}+(2,0,...,0)_{so}\,\,,\\
(2,0,\ldots,0,0)_{su} & \rightarrow & (2,0,\ldots,0)_{so}+(0,0,\ldots,0)_{so}\,\,,\end{eqnarray*}
where $(1,0,\ldots,0,0)_{su}$ is the representation $2m$ of $su(2m)$,
$(1,0,\ldots,0,1)_{su}$ corresponds to the adjoint representation
of $su(2m),$ and the massive gauge fields are in the representation
$(2,0,...,0)_{so}$ of $so(2m)$. Since the $\mathbb{Z}_{2}$ monopoles
can also be associated to roots of the {}``broken'' generators of
$su(2m)$, we could think to associate these monopoles with particles
in the adjoint representation, similarly to the Montonen-Olive case.
Moreover, like for the fundamental BPS $\mathbb{Z}_{2}$ monopoles,
these massive gauge particles have the same mass equal to \cite{KL2010}
\[
m_{W}=ev,\]
for the symmetry breaking caused by the vacuum configuration (\ref{eq:3.21}).
Furthermore, in the ${\cal N}=2'$ SCFT in the non-Abelian Coulomb
phase, they are in a long supermultiplet like the $\mathbb{Z}_{2}$
monopoles, as we explained in the last section. On the other, hand
the massive gauge particles are associated to the weights of $(2,0,...,0)_{so}$
which are not in the coset $\lambda_{1}+\Lambda_{r}(Spin(2m)),$ where
the magnetic weights of the fundamental $\mathbb{Z}_{2}$ monopoles
are in the dual theory. However, in principle the multiplet of the
$\mathbb{Z}_{2}$ monopoles may change at the quantum level.

If by semiclassical analysis the fundamental monopoles remain in the
$2m$ representation, they cannot be dual to gauge particles but,
if we consider a different supersymmetric theory, from the above branchings,
the $\mathbb{Z}_{2}$ monopoles could be dual to particles in the
$2m$ of $su(2m),$ since their weights with respect to the unbroken
$so(2m)$ are exactly equal to the magnetic weights of the fundamental
monopoles. If we consider that $\mathbb{Z}_{2}$ monopoles are dual
to particles in a supermultiplet containing spinors, and if the masses
of these spinors are due to the vacuum solution of the scalar $\phi\in2m\otimes2m,$
we should consider chiral spinors $\psi_{L}\in2m$, $\psi_{R}\in\overline{2m}$
(and therefore $\bar{\psi_{R}}\in2m$ ) of $su(2m)$. Then, if \begin{eqnarray*}
\phi & = & \sum_{p,q=1}^{2m}\phi_{pq}\left|e_{p}\right\rangle \otimes\left|e_{q}\right\rangle ,\\
\psi_{L} & = & \sum_{p=1}^{2m}\psi_{Lp}\left|e_{p}\right\rangle ,\\
\bar{\psi}_{R} & = & \sum_{p=1}^{2m}\bar{\psi}_{Rp}\left|e_{p}\right\rangle .\end{eqnarray*}
the theory can have the Yukawa term \[
\lambda'\left(\bar{\psi}_{Lp}\psi_{Rq}\phi_{pq}+H.c.\right)\]
where $\lambda'$ is a coupling constant in the dual theory. With
the vacuum solution (\ref{eq:3.21}) with constant $v'$, all spinors
become massive with the same classical mass equal to \[
m_{\psi}=\lambda'v'.\]
If we consider $\lambda'=4\pi/e$ and $v'=v$, we obtain exactly the
classical masses of the fundamental BPS $\mathbb{Z}_{2}$ monopoles
of the original theory. In this case, the fundamental BPS $\mathbb{Z}_{2}$
monopoles (with $q=1$) could be dual to $\psi_{L}\in2m$ and their
antimonopoles (with $q=-1$) to $\psi_{R}\in\overline{2m}$. That
would be consistent with the property discussed at the end of Sec.
3 that the set of fundamental $\mathbb{Z}_{2}$ monopoles with $q=1,$
together with the set of fundamental monopoles with $q=-1,$ has a
behavior analogous to particles in two complex conjugated representations.
It is interesting that the symmetry breaking by a scalar $\phi$ in
the representation $2m\otimes2m$ which gives rise to the $\mathbb{Z}_{2}$
monopoles also gives mass to spinors in the $2m$ of $SU(2m)$. Note
that a theory with this field content cannot be embedded in the ${\cal N}=2$
super Yang-Mills theory, like the one discussed in the previous section,
since $\psi_{R}$ and $\psi_{L}$ are in different representations.
However, it can be embedded for example in a ${\cal N}=1$ super Yang-Mills
theory. In \cite{Strassler} it is also considered a duality between
$\mathbb{Z}_{2}$ monopoles and spinors in ${\cal N}=1$ super Yang-Mills
theory but in this case if one theory has gauge group $SU(N)$ with
$N_{F}$ flavors, the dual theory would have gauge group $SU(N_{F}-N+4),$
similarly to Seiberg duality\cite{Seiberg duality}.

\section{Conclusions}

In this work we constructed explicitly BPS $\mathbb{Z}_{2}$ monopole
solutions in theories with the gauge group $SU(n)$ broken to $Spin(n)/\mathbb{Z}_{2}$
using two different vacua of a scalar field in the $n\otimes n$ representation.
Each $\mathbb{Z}_{2}$ monopole is associated to a $su(2)$ subalgebra
and an integer $q$. The magnetic weights of the so-called fundamental
$\mathbb{Z}_{2}$ monopoles correspond to the weights of the defining
representation of the dual algebra $so(n)^{\vee}$. We calculated
the masses for the BPS monopoles and obtained that the fundamental
BPS $\mathbb{Z}_{2}$ monopoles have the same masses and the masses
are equal to $4\pi v/e$, where $v$ is the norm of the Higgs vacuum.
On the other hand, the masses of the nonfundamental $\mathbb{Z}_{2}$
monopoles are the sum of the masses of the constituent fundamental
monopoles. This result is consistent with the interpretation that
the nonfundamental monopoles should be multimonopoles composed of
noninteracting fundamental monopoles, in the BPS case, similarly to
what happens for the $\mathbb{Z}$ monopoles. We showed that the potential
of ${\cal N}=2$ $SU(n)$ super Yang-Mills theories with a hypermultiplet
in the $n\otimes n$ representation, which has a vanishing $\beta$
function, accepts the vacua solutions which break the gauge group
$SU(n)$ to $Spin(n)/\mathbb{Z}_{2}$ . These vacua correspond to
certain points of the Higgs branch where the $\mathbb{Z}_{2}$ monopoles
can exist. It is interesting to note that the BPS equations for the
$\mathbb{Z}_{2}$ monopoles do not result on vanishing of any supercharges.
Therefore, even the BPS $\mathbb{Z}_{2}$ monopoles satisfying first
order BPS equations are in long ${\cal N}=2$ massive supermultiplets,
like the massive gauge fields in this theory. We discussed some possible
dualities the $\mathbb{Z}_{2}$ monopoles may satisfy.

\subsection*{Acknowledgments}

M.A.C.K. wishes to thank L.A. Ferreira, T. Hollowood and N. Seiberg
for many useful discussions. P.J.L. is grateful to CAPES for financial
support.

\end{document}